\documentclass{article}

\usepackage[nonatbib, final]{tackling_climate_workshop_style}

\pdfoutput=1

\usepackage[utf8]{inputenc} 
\usepackage[T1]{fontenc}    
\usepackage{hyperref}       
\usepackage{url}            
\usepackage{booktabs}       
\usepackage{amsfonts}       
\usepackage{nicefrac}       
\usepackage{microtype}      

\usepackage[numbers]{natbib}
\usepackage{graphicx}
\usepackage{amsmath}
\usepackage{subcaption}
\usepackage{{siunitx}} 
\usepackage{wrapfig}

\usepackage{comment}
\usepackage{color}
\newcommand{\tak}[1]{{\textcolor{blue}{ Tak: #1 }}}

\setcitestyle{numbers}

\title{A 3D super-resolution of wind fields via physics-informed pixel-wise self-attention generative adversarial network}

\author{%
  Takuya Kurihana \\
  Department of Computer Science\\
  University of Chicago\\
  \texttt{tkurihana@uchicago.edu} \\
\And
    Kyongmin Yeo \\
    IBM Research \\
    \texttt{kyeo@us.ibm.com} \\
\And
    Daniela Szwarcman \\
    IBM Research \\
    \texttt{daniela.szw@ibm.com} \\
\And
    Bruce Elmegreen \\
    IBM Research \\
    \texttt{bge@us.ibm.com} \\
\And 
    Karthik Mukkavilli \\
    IBM Research \\
    \texttt{karthik.mukkavilli@ibm.com} \\
\And
    Johannes Schmude \\
    IBM Research \\
    \texttt{johannes.schmude@ibm.com} \\
\And 
    Levente Klein \\
    IBM Research \\
    \texttt{kleinl@us.ibm.com} 
}

\begin{document}

\maketitle

\begin{abstract}
To mitigate global warming, greenhouse gas sources need to be resolved at a high spatial resolution and monitored in time to ensure the reduction and ultimately elimination of the pollution source. 
However, the complexity of computation in resolving high-resolution wind fields left the simulations impractical to test different time lengths and model configurations. 
This study presents a preliminary development of a physics-informed super-resolution (SR) generative adversarial network (GAN) that super-resolves the three-dimensional (3D) low-resolution wind fields by upscaling $\times$9 times. 
We develop a pixel-wise self-attention (PWA) module that learns 3D weather dynamics via a self-attention computation followed by a 2D convolution. We also employ a loss term that regularizes the self-attention map during pretraining, capturing the vertical convection process from input wind data. 
The new PWA SR-GAN shows the high-fidelity super-resolved 3D wind data, learns a wind structure at the high-frequency domain, and reduces the computational cost of a high-resolution wind simulation by $\times$ 89.7 times.   
\end{abstract}

\section{Introduction}\label{sec:intro}
Accurate tracing and monitoring of greenhouse gas (GHG) sources is a key measurement to take action for tackling the mitigation of global warming issues~\cite{defries2007earth}.
The higher spatial resolution of wind simulation enables precise tracking of GHG emissions from potential sources, and helps decision-making in various fields such as policymakers, agriculture, and renewable energy. 
However, high-resolution simulation is not always available due to the demands of computational resources. 

Artificial Intelligence (AI), including deep neural networks (DNNs), can reduce the computational cost by 
upscaling the wind fields from low-resolution (LR) to high-resolution (HR) data.
Super-resolution (SR) is one of the solutions to achieve the goal: conventional SR techniques rely on convolutional neural networks (CNNs) to reproduce realistic HR wind fields~\cite{stengel2020adversarial, xie2018tempogan,kurinchi2021wisosuper, liu2022spatial, km22sr2d, annau2023algorithmic}.  
However, these CNN-based approaches either perform on 2D data or fall short of capturing the 3D dynamics in weather systems because a convolutional kernel truncates the association between vertical layers, limiting to learning of convection and diurnal cycles induced by incoming solar radiations. 
Another issue is that widely used weather simulations employ a non-uniform grid in a vertical axis for computational stability.  
The difference in height between a pair of two adjacent vertical layers becomes larger at higher altitudes. Thus, CNN is not sufficient to capture physics at the same vertical scaling.

To address these issues, we develop a prototype of a novel architecture of physics-informed neural networks that learn multi-scale spatial dynamics. Our goal is to develop a physics-informed neural network that super-resolves 900~m resolution of three-dimensional wind data into 100 m scale high-resolution wind data and captures three-dimensional dynamics in weather systems. 

\section{3D super-resolution of wind fields}We develop a \emph{pixel-wise self-attention network} (PWA) to learn the three-dimensional dynamics of weather simulations. Using the neural network as a generator, we train a generative network for super-resolving wind velocity fields. 

\subsection{WRF model dataset}\label{sec:wrf}
The Weather Research \& Forecasting Model (WRF)~\cite{skamarock2008description} has been widely used for simulating weather systems from mesoscale to turbulence scales. Large-eddy simulation (LES) is often nested into the WRF framework to allow simulating a convective process to capture a more realistic turbulence structure~\cite{de2013entrainment}. 
Outputs from WRF-LES used for this study are nested in 900~m, 300~m, and 100~m horizontal resolution. The outer 900~m and inner 100~m simulations differ in their model physics and our ultimate goal is to super-resolve 900~m LR data to 100~m HR data. This study, however, focuses on synthesizing the 900~m LR data by spatially averaging 100~m HR data as a preliminary step. 
Our training and testing data are produced by a WRF nested LES at 100~m horizontal scale and 59 vertical layers, generating outputs every five minutes from 00 UTC to 23:59 UTC. The initialization time window is set for two hours starting at 22:00 UTC a day before. We sample training and testing data randomly from one-month simulation outputs from September 2019. We select three wind velocity fields (U, V, W), all at 100 m spatial resolution over 40 km $\times$ 40 km simulation domain, and use only 8 layers from the surface level. 
We then extract \num{22500} training and \num{2500} testing data, as the smaller geographical images, each of which is 126 pixels $\times$ 126 pixels ($\sim$ 12.6 km $\times$ 12.6 km) for high-resolution (HR) data.
To synthesize low-resolution (LR) data, we average every $9 \times 9$ pixels, giving 900 m LR data.

\subsection{Physics-informed pixel-wise self-attention}\label{sec:self-attention}

\begin{figure}[ht!]
\centering
\begin{subfigure}{0.54\textwidth}
    \centering    
    \includegraphics[width=1\textwidth]{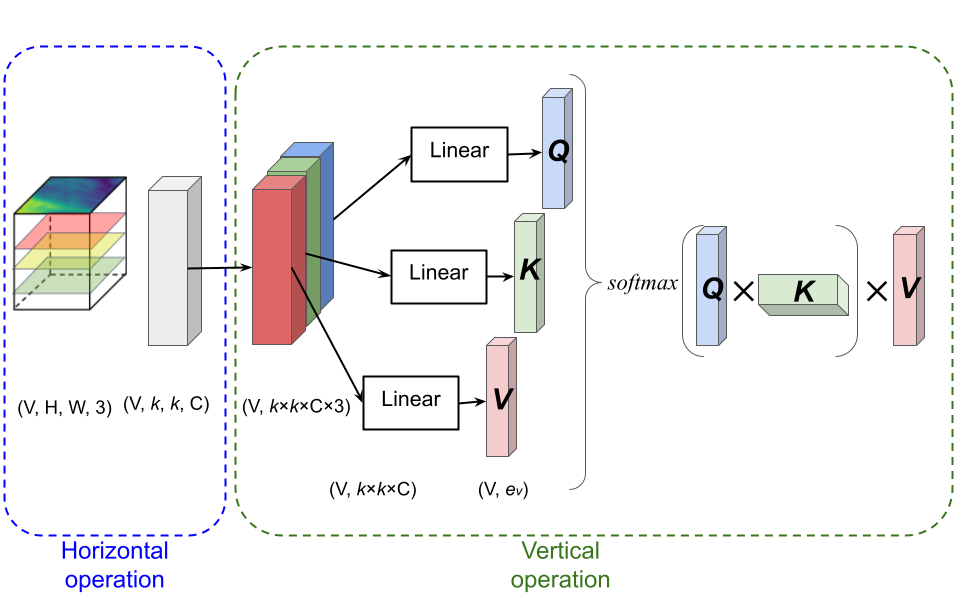}
    \subcaption{Diagram of the pixel-wise self-attention module. }
    \label{fig:pwa-module}%
\end{subfigure}
\begin{subfigure}{0.45\textwidth}
    \centering    
    \includegraphics[trim={1 1 1 1},clip, width=\textwidth]{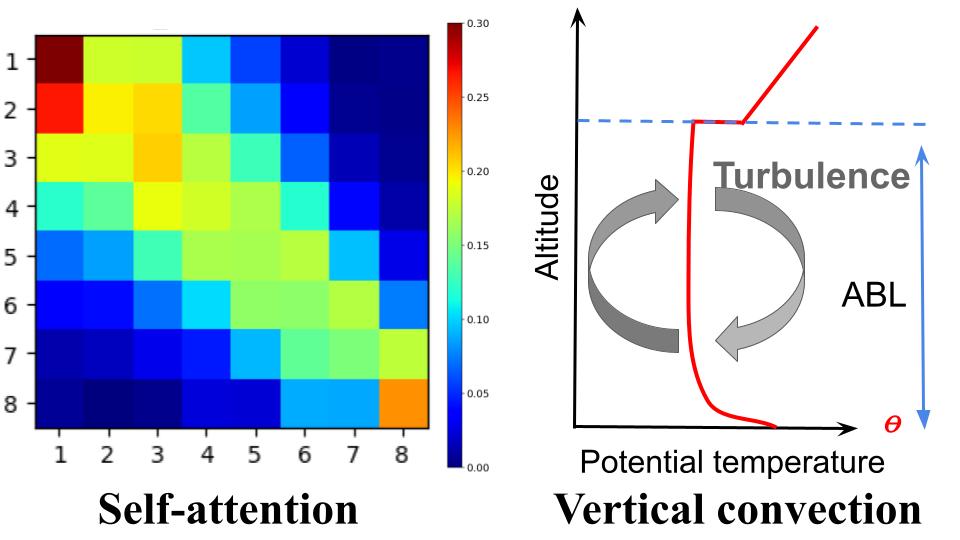}
    \subcaption{Physical interpretation. }
    \label{fig:pwa-module-interp}%
\end{subfigure}
\caption{Illustration of the pixel-wise self-attention module: (a) workflow of PWA module. (b) an example 8 $\times$ 8 self-attention map (\textbf{left}) shows strong (bright colors) and weak (dark colors) signals between adjacent layers, associating convection in weather systems (\textbf{right}).  }\label{fig:pwa-concept}
\end{figure}

The fluctuation in the atmospheric boundary layer (ABL) height during both day and night significantly impacts the development of vertical convection intensity, which varies across different altitudes. 
A standard convolutional filter may truncate these signals and be suboptimal to the non-uniform vertical grid system in numerical models for capturing information at the same scale.
We introduce a self-attention computation~\cite{vaswani2017attention} at each grid column to better embed nonlinear association between vertical layers into networks: 
self-attention inherently computes attention scores for each element in sequences, enabling the representation of signal associations between a given vertical layer and others. 
This versatility extends beyond the constrains of a convolutional kernel that limits associations solely to neighboring layers above and below.

\autoref{fig:pwa-concept} shows the concept of the newly developing pixel-wise self-attention (PWA) module at a high level. Suppose we have an input image $X$, a 5D Tensor $(N, C, V, H, W)$ where N, C, V, H, and W depict the size of the mini-batch, the number of variables (i.e., U, V, and W winds), the number of vertical layers, height, and width. PWA module is composed of two parts: \emph{horizontal} and \emph{vertical} operation.
For the horizontal operation, we apply the 2D convolutional operation to each image of $(V, H, W)$ by $1 \times 3 \times 3$ convolution kernel.
For the vertical operation, we first convert the image shape from $(V, H, W, C)$ to $(V,  H \times W \times 3\cdot C)$ with a linear transformation to create three matrices such that query $Q \in \mathbb{R}^{V \times e_v}$, key $K \in \mathbb{R}^{ V\times e_v}$, and value $ V \in \mathbb{R}^{ V\times e_v}$. 
We then calculate a self-attention computation by a simple scaled dot product in~\autoref{eq:self-attention1} as follows;
\begin{wrapfigure}[21]{r}{0.38\textwidth}
    \centering
    \vspace{-5mm}
    \includegraphics[trim={25 5 25 5}, clip, width=.65\textwidth, angle =-90 ]{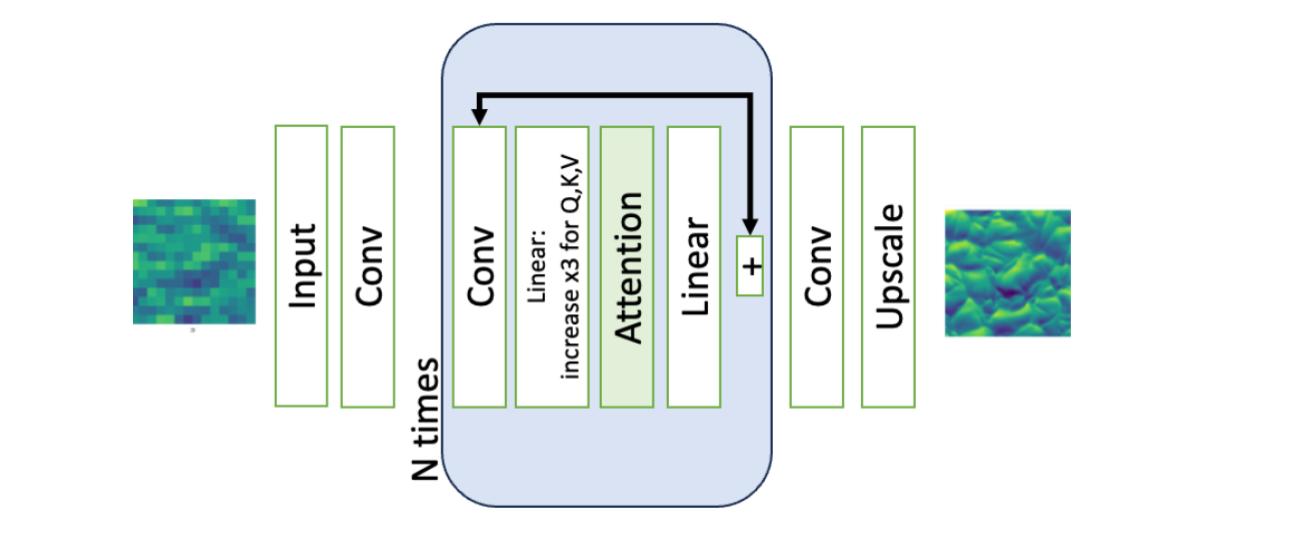}
    \vspace{-15mm}
    \caption{Diagram of the architecture of PWA-SR network. We nest the PWA module (highlighted by light blue color) by three times based on our hyperparameter search. }
    \label{fig:network}
\end{wrapfigure}
\begin{equation}\label{eq:self-attention1}
    \mathrm{Attn} = \underbrace{\mathrm{softmax} \left( \frac{Q \cdot K^T}{ \sqrt{e_v} } \right) }_{M} \cdot V, 
\end{equation}
where $e_v$ is a dimension of intermediate representation and we depict 64 for the dimension.
The learned self-attention map $M$ shown in~\autoref{fig:pwa-module-interp} shows strong signals in diagonal and near-diagonal elements, indicating 
that adjacent layers have a stronger effect on each other.
In particular, near-surface layers (i.e., left top corner of $M$) exhibit strong attention in wider elements, and this is associated with daytime convection within the ABL. 

Effective learning is achieved through the sparsity of weak signals as well as emphasizing key elements in the self-attention map $M$ depicted in Figure~\ref{fig:pwa-module-interp}.   
To do so, we add the regularization of $M$ as one of the loss terms as follows: 
\begin{eqnarray}\label{eq:self-attention2}
    R_i  & = & \sum_{j=1}^{V} M_{ij}^2  , \\
    R(M) & = & \sum_{i=1}^{V} \frac{1}{R_i},
\end{eqnarray}
where $M_{ij} = \mathrm{softmax} ( Q K^T  / \sqrt{e_v} )$ and $R$ depicts a regularization term. 
The minimization of $R(M)$ in Equation~\ref{eq:self-attention2} indicates the elements of strong attention get larger scores as well as weaker attention reduces the scores. 
One limitation of this approach is that the range is bounded in $[0,1]$ at each row. To have more attention to strong signals, and in opposite, suppress the weak ones, we introduce a re-scaling scheme to the self-attention map. 
Simply, we have a trainable matrix $\Delta$, adjusting the value of self-attention map $M$ by $M^{\mathrm{rescale}} = \Delta \cdot M $. 
Note that we apply the self-attention map regularization scheme to pre-rescaled $M$.

\subsection{PWA SR-GAN: Pixel-wise self-attention super-resolution generative adversarial network}\label{sec:gan}

\autoref{fig:network} illustrates the architecture of our super-resolution network inspired by SR-GAN~\cite{ledig2017photo}. 
Due to the nature of convolutional filters that pass low-frequency modes, the results of neural networks have smoothed structures. To address this, we additionally include Sobel filter~\cite{zhang09sobel-apply} by computing image gradients in horizontal and vertical directions respectively due to the different spatial scales.
Overall, we combine multiple loss terms as follows;
\begin{equation}\label{eq:regularized_gradient_uvw_loss}
    \mathcal{L} = \mathcal{L}_{\mathrm{content}} + \lambda_{R} | R(M) | +  \underbrace{ \lambda_{G} \sum_{r\in\{u,v\}} | dx/dr - d\hat{x}/dr  | + \lambda_{G_v} |dx/dz - d\hat{x}/dz| }_{\text{Image gradient}},
\end{equation}
where $\mathcal{L}_\mathrm{content} $ represents mean squared error $\sum_{ x_{LR} \in X_{LR} , x_{HR} \in X_{HR} }  \left( G(x_{LR}) - x_{HR} \right)^2$; $G(\cdot)$ depicts the outputs of our neural network. We have three different weight coefficients: $\lambda_R$, $\lambda_{G}$, and $\lambda_{G_v}$ balances self-attention regularization, horizontal gradient, and vertical gradient terms. 
We use $(\lambda_R,\lambda_{G}, \lambda_{G_v}) = (0.01, 50, 100)$ in this work. The training uses Adam optimizer~\cite{kingma2014adam} with the learning rate at $1e-3$ on an A-100 GPU for 200 epochs.

CNN-based neural networks are known for filtering low-frequency data in input images and removing high-frequency information~\cite{magid2021dynamic}. 
Generative Adversarial Networks (GANs)~\cite{goodfellow2020generative} are our solution to incorporate them into super-resolved images because our training input data $x_{\mathrm{LR}}$ has already smoothed out high-frequency information in high-resolution image $x_{\mathrm{HR}}$ through averaging the data over $9 \times 9$ pixels. Thus, it is challenging to learn the high-frequency data only from low-resolution training inputs. 
The generator loss in a GAN is typically defined as the negative log-likelihood of the discriminator's output when the generator tries to generate realistic data. The loss term $\mathcal{L}_{\mathrm{adversarial}} $ is often represented as:
\begin{equation}
\mathcal{L}_{\mathrm{adversarial}} = -\mathbb{E}_{\mathbf{x^{\prime}} \sim p(\mathbf{x_{\mathrm{LR}}})}[\log(D(G(\mathbf{x^{\prime}})))],    
\end{equation}
where $\mathbf{x^\prime}$ is a set of low-resolution wind data sampled from a distribution of low-resolution images; 
$G(\mathbf{x^\prime})$ is a super-resolved data by the generator;
$D(G(\mathbf{x}))$ is a discriminator's output for high-resolution data $\mathbf{x}$; and 
$p(\mathbf{x}^\prime)$ is a prior distribution of $\mathbf{x}^\prime$. 

After pre-training PWA-SR network with~\autoref{eq:regularized_gradient_uvw_loss}, we modify the loss function for training a generator by adding an adversarial term into the combined loss function;
\begin{equation}\label{eq:generator-loss}
    \mathcal{L} = \mathcal{L}_{\mathrm{content}} + \lambda_{G} \mathcal{L}_{\mathrm{horizontal}} + \lambda_{G_v} \mathcal{L}_{\mathrm{vertical}} + \lambda_a \mathcal{L}_{\mathrm{adversarial}}.
\end{equation}
For discriminator, we follow the same architecture based on Ledig et al.~\cite{ledig2017photo}. 
See Ledig et al.~\cite{ledig2017photo} for further details on the training of GAN. 

\section{Super-resolution of 3D wind data}\label{sec:results}

\begin{figure}[ht!]
    \centering
    \includegraphics[width=1\textwidth]{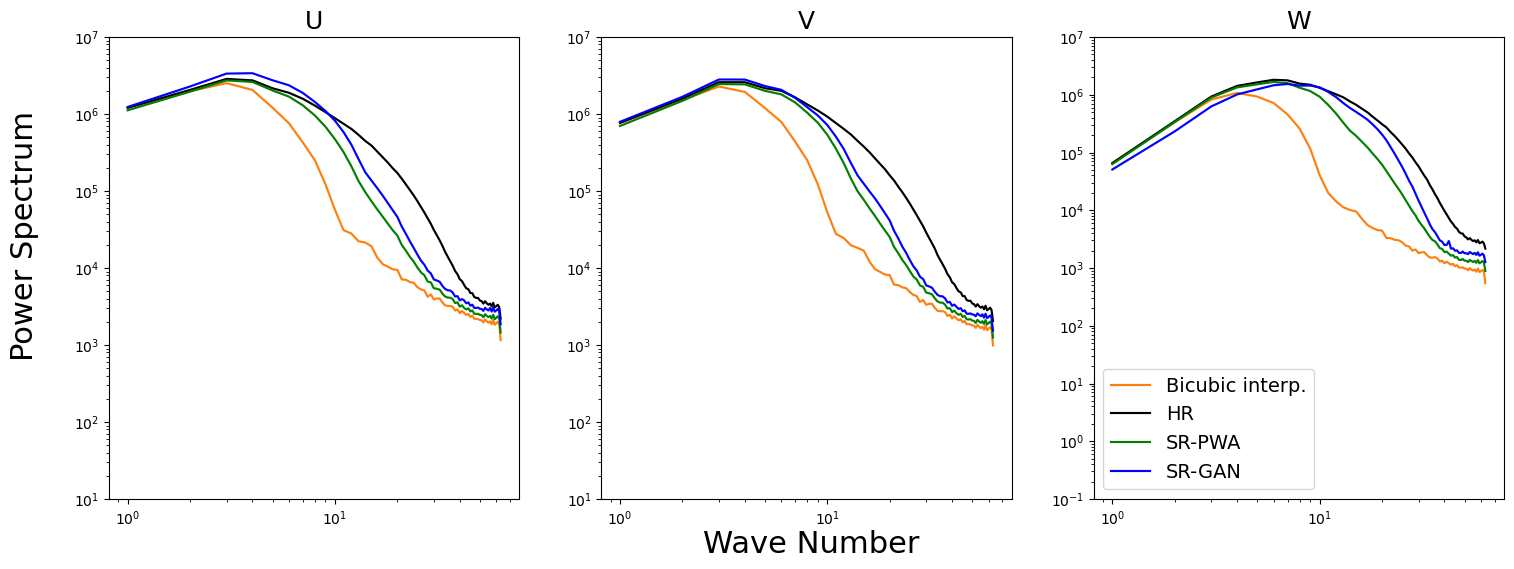}
    \caption{Plots of power spectrum analysis for bicubic interpolation, pixel-wise self-attention SR network (SR-PWA), and pixel-wise self-attention SR-GAN (SR-GAN) as a comparison to HR images. SR-GAN outperforms the SR-PWA (green line) for learning high-frequency domains, especially in vertical wind W }
    \label{fig:power}
\end{figure}

We first investigate whether PWA SR-GAN (hereafter SR-GAN) can generate wind filed data that closely approximates the ground-truth HR image in frequency space. 
We apply the fast Fourier transform to our \num{2500} test samples for calculating the power spectrum of images instead of mean squared errors (MSEs) because MSEs do not always evaluate if SR models restore high-frequency mode. 
That is, the test measures how well the model learns the high-frequency mode of wind data through GAN training.
\autoref{fig:power} shows the results of the power spectrum analysis as a function of wave numbers for U, V, and W wind data respectively. The results compare the power spectrum among different approaches: HR data, bicubic interpolation, SR-PWA, and SR-GAN. The fidelity of capturing realistic wind structures improves as their results of the power spectrum align more closely with HR data (black line).
Note that a conventional residual CNN yields a result similar to that of SR-PWA (not reported in this paper). 
We see that the SR-GAN approach (blue line) significantly outperforms the SR-PWA (green line) for learning high-frequency domains, especially in vertical wind W. However, there is a performance trade-off in low to medium-wave numbers when the learning high-frequency domain achieves well. 

\begin{figure}[ht!]
    \centering
    \includegraphics[width=1\textwidth]{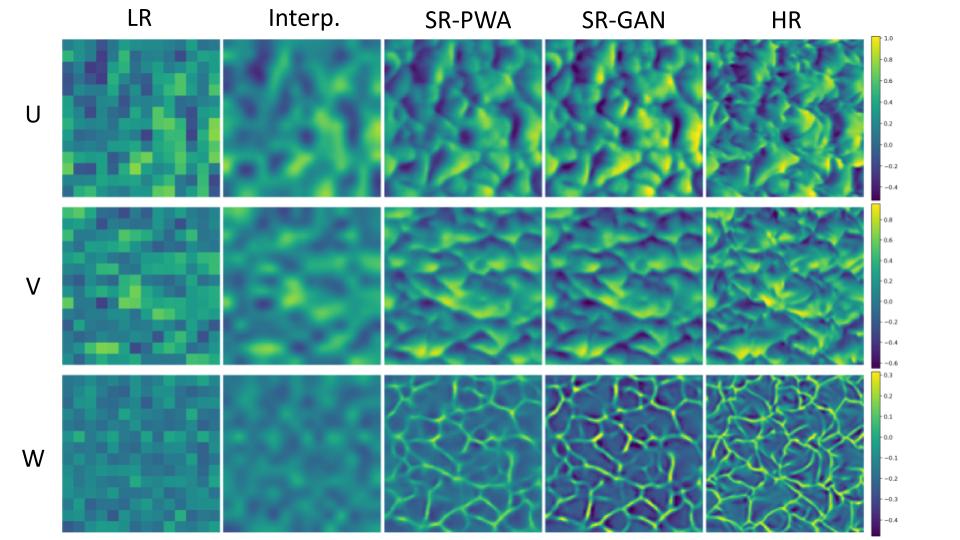}
    \caption{Comparison of qualitative test results based on an example snapshot at the first layer. Each row shows the super-resolved or raw images from U, V, or W wind components.}
    \label{fig:example}
\end{figure}
As part of a qualitative evaluation, we present a snapshot example of wind data from the first layer in \autoref{fig:example}, illustrating the fidelity of super-resolved wind images as a comparison between LR data and HR data.
SR-PWA restores major wind structures and the velocity intensities.
SR-GAN excels in generating finer wind structures, as evidenced by the results depicted in~\autoref{fig:power}. 
Additionally, we observe that our SR-GAN decreases computational costs ($\sim$ CPU time $\times$ number of cores) 
by $\times$89.7 times compared with the original WRF-LES.   
Overall, the results indicate that PWA SR-GAN generates realistic 3D high-dimensional wind fields by lower computing power and resources, enabling simulating advection and diffusion of GHG gas solely using low-resolution WRF simulation outputs.

\section{Summary}\label{sec:summary}
In summary, we show our preliminary investigation of the super-resolved 3D wind structures based on a newly developing SR network that utilizes a self-attention network and a generative model.
It enables to incorporating of 3D dynamics of weather systems that are essential to reconstructing physically representative 3D wind fields, and then achieve to generate high-fidelity outputs. 
This work is a preliminary step toward building a tracing methodology that is capable of simulating the trajectory of GHGs combined with limited observation and reduces the computational overhead.  
Further algorithmic advancements should improve the accuracy of the methodology. 

\begin{ack}
Supports for this work come from summer internship program at IBM Research, Yorktown Heights. The authors thank NeurIPS 2023 Tackling Climate Change with Machine Learning workshop committees for having opportunities to present our research outcomes.
\end{ack}

\medskip
\bibliographystyle{plain} 
\bibliography{references}

\end{document}